\documentstyle[11pt,mrs2001,epsfig]{article}

\begin{document}
\title{Spectral Reduction Software for the DEEP2 Redshift Survey}
\author{C. Marinoni, M. Davis, A. Coil}
\affil{Department of Astronomy, University of California at Berkeley, Berkeley, CA, 94720, USA}
\author{D. Finkbeiner}
\affil{Department od Astrophysics, Princeton University, Princeton, NJ 08544, USA}

\begin{abstract} The DEEP2/DEIMOS redshift survey, 
which will begin observing in the Spring of 2002,
will gather high quality spectra on $\sim 60000$ galaxies 
in order to study the evolution of the properties and large 
scale clustering of galaxies at $z \sim 1$.

The data rate
from DEIMOS  will be in excess of 1 Gbyte/hour, and it is therefore
imperative to  employ completely  automated data  reduction techniques to manage the
analysis.  We here describe aspects of our data pipeline, which will make
extensive use of B-splines for the sky-subtraction stage and for the
combination of multiple frames.  
\end{abstract}

\section{Spectral Reduction Strategy}
The Keck II Deep Imaging Multi-Object Spectrograph (DEIMOS) 
is intended for imaging and multi-slit spectroscopy
over a field of view that is approximately a rectangle of size 16' by 5'
and over the wavelength range 0.42-1 $\mu m$.
With its focal plane mask allowing slitlets spectroscopy for $\sim$ 100-150 
objects, 
DEIMOS is optimally designed for large surveys of faint objects.
The initial use of DEIMOS will be largely to undertake a massive redshift
survey of $\sim$ 60,000 $z \approx 1$ galaxies, the DEEP2 survey \cite{1}.

The technical complexity of the reductions and the shear volume of data  
force us to a rather new approach to spectral data reduction. 
Working closely with the SDSS team, we are   developing a dedicated, 
fully automatic 
data reduction package  based on the IDL codes of   
Schlegel, Burles, and Finkbeiner. 

\subsection{Spectral Tracing}
The DEIMOS mosaic imager is an 8K by 8K pixel camera composed of 8 
individually mounted 2K by 4K CCDs manufactured by MIT Lincoln Laboratory.
The pixel size is 15 $\mu$m. The DEIMOS CCDs are thick, high
resistivity devices, with enhanced QE in the near-IR and much reduced fringing
compared to thinner chips.   In contrast to the VLT/VIMOS design, 
DEIMOS will have smaller multiplexing but many more spectral pixels per target,
and there will not be multiple objects within a given row of data.
The large number of pixels in the dispersion direction (8K)
allows high resolution with substantial spectral range, i.e. a wide redshift
interval coverage,
while the spatial extension allow us to cover 16' of sky and a large 
multiplexing capability.

The spectrum of each object will be dispersed across
2 CCDs separated by a gap whose size is a polynomial function of the position 
along the spatial direction. We have developed a series of fast algorithms 
to trace the spectra across the 2 CCDs, align and rectify them and 
fit the gap size with sub pixel precision at each trace position.  Based on the
traces of the edges of the
 curved slitlet spectra over the CCD frame, the data for individual slitlets
is shifted
by whole pixels in the spatial direction to produce rectangular spectra.  
We next perform a  non linear wavelength solution fit 
to calibrating arc lamps using the formula
\[\lambda(x)=a_j \cdot L^j \big[x+G(y)\theta \big(\frac{x}{4096+G(y)}\big)\big]\]
where $L^j$(u) are Legendre polynomials of order j, x is the pixel position along the
dispersion direction, G(y) is the unknown gap at position y along the spatial direction 
and $\theta$ is the Heaviside theta function. A standard Hg-Ne-Ar lamp,
dispersed by a 900 l/mm grating  can be fit with a scatter of 0.01 $\AA$ and
gives the functional form of gap function G.  Since the gaps will not change
with time, the information for G(y) will be stored in a table rather than fit
from individual frames.

\begin{figure}[t]
\plottwo{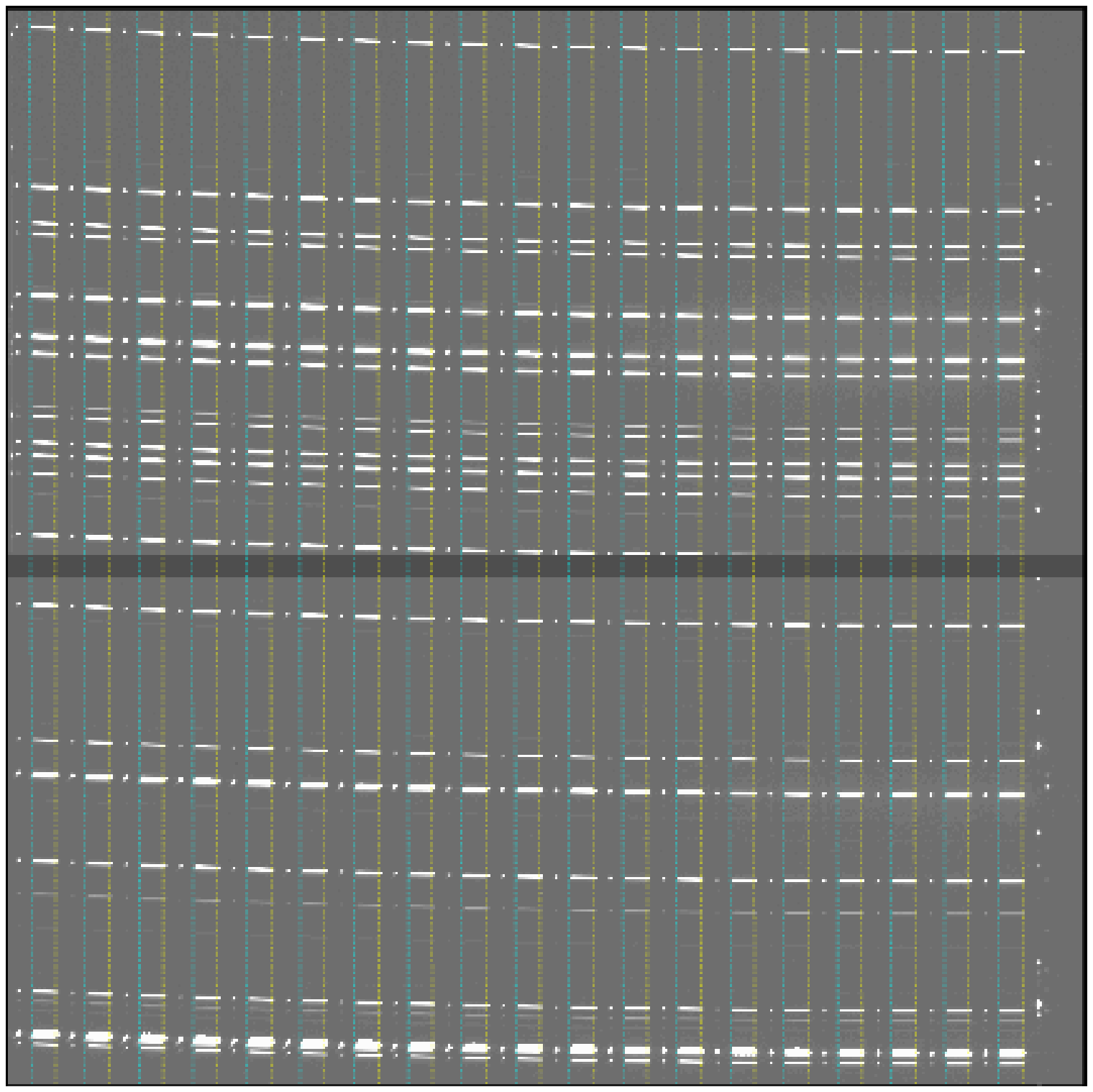}{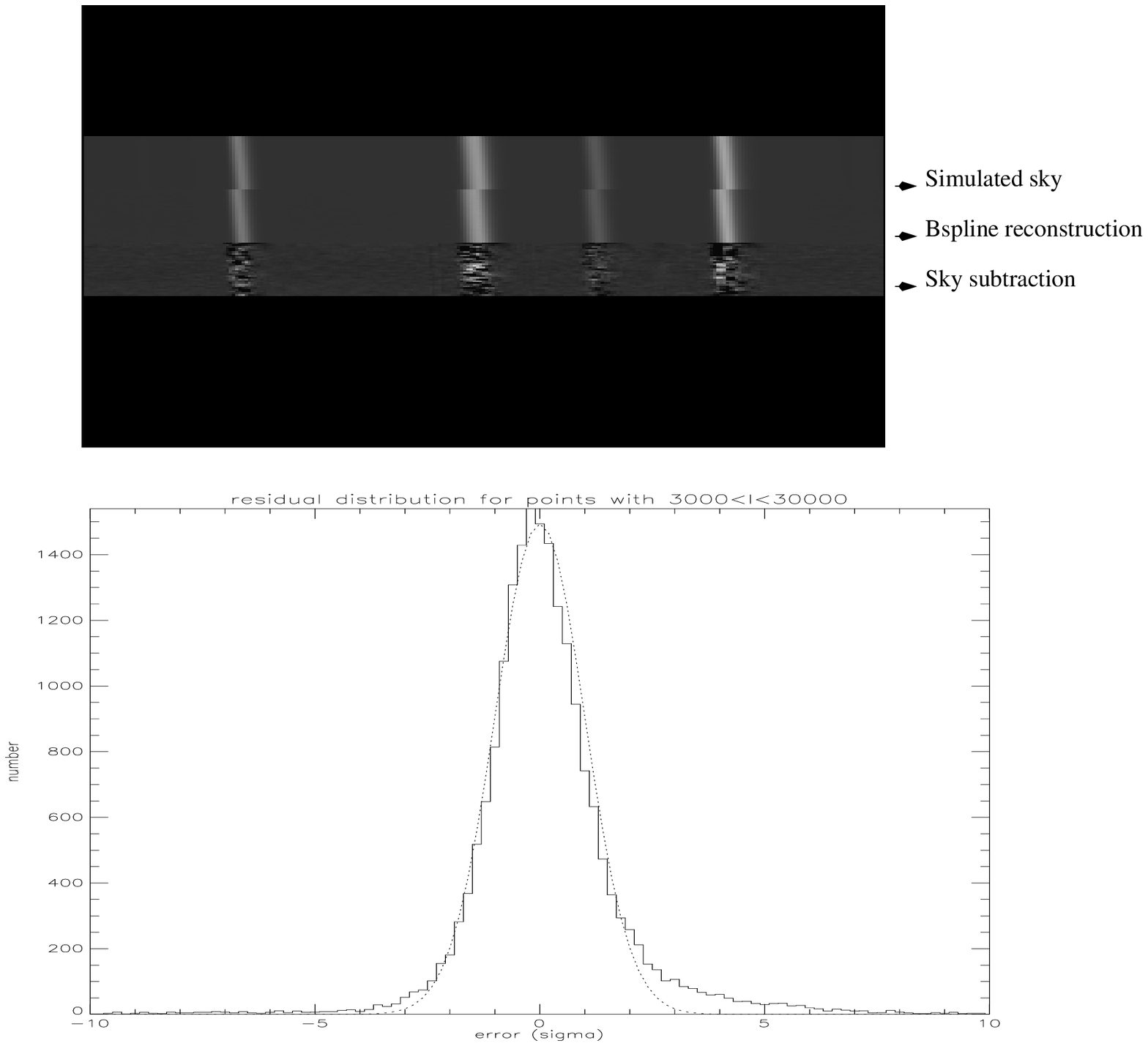}
\caption{{\em Left:} Calibrating arc lines traces falling on 2 different DEIMOS CCDs.  
Spectral tracing and rectification is performed using flat fields
exposures and no data rebinning. Note that the traces are curved in the CCD plane 
{\em Right:} A small section of an extracted 2-d spectrum of an
arc lamp, showing the expected tilt in the spectral lines.
Residuals obtained after subtracting the sky spectrum from its model are 
plotted. The final frame is shown to be  essentially Poisson noise limited.}  
\end{figure}

\subsection{Sky Subtraction}
Studies of faint, high-redshift galaxies rely on our ability to remove very 
strong night sky airglow lines which overlie the
spectra of these faint objects. This is crucial at high redshifts, since the 
intensity and frequency of OH night sky lines 
increases in the near-IR and can readily swamp emission lines from
faint target galaxies, such as a redshifted OII doublet (3727 $\AA$).

The DEIMOS camera coupled with the 900 l/mm grating and slits of width 0.75''
will allow us to observe in a high resolution operational mode, 
$R \equiv \lambda / \Delta\lambda =3700$. This planned 
spectral resolution  (80km/s)
is needed  for resolving the OII doublet (giving confidence to the redshift
determination even if no other features are observed), 
minimizing sky lines spectral contamination, and 
optimizing atmospheric subtraction.   With this high spectral purity, we
intend to use tilted slits to follow the major axis and to map rotation
curves of many of the target 
galaxies, thus leading to seriously tilted night sky lines in the 2-d image.

Sky  subtraction can be a very critical step in the reduction of long slit 
spectra with the main problems
being uncorrected fringing and the curvature of the lines along the slit. 
Although one may intuitively tend to subtract the sky spectrum still in pixel space in order to avoid the problems
inherent to non-linear rebinning, experience shows that a proper wavelength calibration can remove the curvature of
the sky lines to a high degree of accuracy (one should aim for 0.1 pixels rms)
The standard approach to sky subtraction consists of a  linear interpolation 
of the sky modal values of the two sky frames which immediately bracket 
the object frame. But such an approach requires considerable extra slit-length
for each object, reducing the allowed multiplexing of the targets.  
Furthermore, the analysis of data from tilted slits is greatly complicated
in such a scheme.

An alternative optimal method consists in processing 
background  spectral characteristics 
with  B-spline functions. This is a relatively new 
fitting technique, widely applied in CAD systems in which 
a set of piecewise polynomial curves are used
to approximate continuum data.
Here the key-word is approximation, which means that the B-splines
curves pass close to a set of data points without passing exactly through them.
B-spline curves are represented in parametric notation  by the following 
expansion 
\[P(t)=\sum_i^n N_{i,k} P_i\]

where N is the functional basis and $P_i$ are n data points.
We can interpret it as a normal functional expansion but instead 
of using Legendre or Chebishev orthonormal polynomials we use a set of more flexible
functions such as piecewise cubic splines.
 These functions are uniquely determined (as in a standard 
interpolative scheme) once we specify 
an order of fit k and a set of control points (in the domain of the 
parameter t)  known as knots.

B-splines have many attractive properties
that make them superior to more conventional basis sets in the description 
of  continuum data, the main advantage residing in 
their local nature. Any control point  influences the shape of the curve 
close to it
and, with a suitable spacing between knots, we can control the elasticity of 
the fit 
in such a way to improve the stability of the reconstruction 
in noisy image environments. In this way we can automatically reject 
interlopers, 
image defects or cosmic rays polluting the sky determination.
Moreover the convergence to solution is always guaranteed by the intrinsic 
interpolative nature of the method.   The method requires that we know the
wavelength of every pixel, but there is no rebinning of the sky data, and
tilted slits or tilted spectral lines present no additional complication.
The B-spline fitting can be used to generate a ``supersky'' spectrum which
can be applied to different slitlets.  The same technology is ideal for
the averaging of 1-d extracted spectra, without the need to first remove
 cosmic rays or to align the spectra in wavelengthspace

We simulated the 
performance of the B-spline
 method by using arc lamp spectra to  reproduce the
night sky lines. Even if the real sky subtraction accuracy is difficult 
to gauge
with laboratory images, the residual counts in 
our simulated sky-subtracted spectra (shown in fig. 1) provide a good 
estimate of the goodness of the method.  With a sufficient knot density,
we are able to reproduce the spectrum of the arclamp to high precision,
and we find that the residuals of the spectrum from the B-spline average
are consistent with Poisson fluctuations, even over the regions of nearly
saturated emission lines, as shown in figure 1b.

\subsection{2D Wavelength Solution}
Thanks to this capability of 
automatically removing the most deviating points
from the underlying signal, the B-spline procedure  can also be
optimally  applied for determining the full wavelength 
calibration of a 2D extracted spectrum. The 2D pixel-to-wavelength mapping
is needed  in order to apply the sky subtraction strategy described in \S 1.2.

To obtain this information we exploit the simple fact that 
the wavelength scale $\lambda=\lambda(x,y_c, G(y_c))$, established for 
the central reference row $y_c$  of the comparison spectra (arc lamp), 
allows us to unambiguously label, with tabulated wavelengths, 
every other peak position ($x_i,y_j$) of the spectra.

We then fit at once the whole comparison spectra (i.e. the data set 
$\lambda=\lambda(x_i,y_j)$) with a 2-dimensional  B-spline 
generating  the full wavelength solution for each pixel in a fast  
way, virtually insensitive to bad pixels or outliers.

\bigskip
\bigskip

The Keck/DEIMOS survey is a collaborative project among astronomers
at UC, Caltech and the Univ. of Hawaii. The team 
intends to share all results with the public
and to put the spectra online in a timely manner. Details of the project 
can be found at the URL {\bf http://astro/berkeley.edu/deep/}

\end{document}